# Search for Antineutrino Charged Current Coherent Pion Production at SciBooNE

Hide-Kazu TANAKA

*Massachusetts Institute of Technology, Department of Physics,*
*77 Massachusetts Avenue, Cambridge, MA 02139-4307*

**Abstract.**
The SciBooNE experiment (Fermilab) recently published results of a search for charged current coherent pion production in neutrino mode: muon neutrinos scattering on carbon, $\nu_\mu{}^{12}\mathrm{C} \rightarrow \mu^{-}{}^{12}\mathrm{C}\pi^{+}$. The results of this study are that no evidence for coherent pion production is observed, and SciBooNE set 90% confidence level upper limits on the cross section ratio of charged current coherent pion production to the total charged current cross section. Recently proposed new coherent pion models predict a production of charged current coherent pion events just below the SciBooNE's upper limit. Motivated by this, we performed a search for charged current coherent pion production using SciBooNE's collected antineutrino data since antineutrino data are expected to be more sensitive to look at coherent pion production than neutrino data.

This paper describes preliminary results of a search for antineutrino charged current coherent pion production at the SciBooNE experiment.

## INTRODUCTION

The unknown neutrino mixing angle $\theta_{13}$ is one of the most important goals in current neutrino experiments. For the next generation of long baseline neutrino oscillation experiments, T2K and NOvA, the precise measurement of neutrino-nucleus cross sections in the few GeV energy range is an essential ingredient in the interpretation of neutrino oscillation signals. Charged current single charged-pion production (CC-1$\pi$) is a dominant background process for $\nu_\mu \rightarrow \nu_X$ oscillation measurements.

Neutrino-induced charged current single charged pion production is dominated by baryonic resonance excitation off a single nucleon bound in a nucleus in the neutrino energy region of a few GeV. The resonance state is followed by its prompt decay into a nucleon and a pion in the final state. The process is written as $\nu_\mu N \rightarrow \mu^- N' \pi^+$ where $N$ and $N'$ are proton or neutron. In addition to this reaction, neutrinos can produce pions by interacting *coherently* with the nucleons forming the target nucleus. The process is expressed as $\nu_\mu A \rightarrow \mu A \pi^+$, where $A$ is a nucleus. Understanding the cross sections of these processes is important to study $\nu_\mu \rightarrow \nu_X$ oscillation ($\nu_\mu$ disappearance) near one GeV.

Coherent pion production in neutrino-nucleus interactions has already been the subject of several experimental campaigns. The neutrino energy range between 2 and 100 GeV has been investigated, including both the charged current and neutral current modes, and using both neutrino and antineutrino probes [1, 2, 3, 4, 5, 6, 7, 8, 9, 10]. The results in the high energy region are well described by the Rein-Sehgal model [11] prediction that is widely used in neutrino experiments today.

Recently, there are two new results on coherent pion production, and they are drawing a lot of attention in the neutrino physics community. The non-existence of charged current coherent pion production in a 1.3 GeV wide-band neutrino beam has been reported by K2K [12], while there exist charged current coherent pion production positive results at higher neutrino energies. On the other hand, evidence for neutral current coherent pion production at a similar neutrino energy has been reported by MiniBooNE [13].

The SciBooNE experiment recently published results [14] of a search for charged current coherent pion production from muon neutrino scattering on carbon (based on the Rein-Sehgal model [11, 15]), with two distinct data samples; averaged energy 1.1 GeV and 2.2 GeV. No evidence for coherent pion production was observed. SciBooNE set 90% confidence level upper limits on the cross section ratio of charged current coherent pion production to the total charged current cross section at $0.67 \times 10^{-2}$ at a mean neutrino energy of 1.1 GeV and $1.36 \times 10^{-2}$ at a mean neutrino energy of 2.2 GeV.

Along with the experimental studies of coherent pion production, new theoretical predictions for coherent pion production also became available. Several theoretical models describing coherent pion production have been proposed, e.g. [16, 17, 18, 19, 22, 21, 22], using different formalisms to describe the relevant physics. These new models predict smaller charged current/neutral current coherent pion production cross sections than the Rein-Sehgal model prediction and predict a charged current coherent pion cross section just below SciBooNE's upper limit.

Because of the connection between neutrino and antineutrino coherent pion production processes in the theoretical models, it will be interesting to repeat this analysis on SciBooNE's already collected antineutrino data. Most models predict a similar absolute cross section for neutrino and antineutrino coherent pion production, which means the ratio of charged current coherent pion events to charged current inclusive events is expected to be larger in antineutrino data because of the reduced total $\bar{\nu}$ charged current event rate. Because of this, the antineutrino search has the potential to be even more sensitive. Motivated by this, SciBooNE has performed a charged current coherent pion production search in antineutrino mode.

## SCIBOONE EXPERIMENT

The SciBooNE experiment [23] is designed to measure neutrino cross sections on carbon in the one GeV region. The experiment collected data from June 2007 to August 2008 with neutrino and antineutrino beams in the FNAL Booster Neutrino Beam line (BNB). The BNB uses a primary proton beam with kinetic energy 8 GeV and a beryllium target placed inside an aluminum horn as a neutrino production target. The SciBooNE detector is located 100 m downstream from the neutrino production target. The neutrino flux in the SciBooNE detector is dominated by muon neutrinos ($\sim 93\%$) in neutrino running mode. The flux-averaged mean neutrino energy is 0.7 GeV. When the horn polarity is reversed, $\pi^-$ are focused and hence a predominantly antineutrino beam is created, with mean energy 0.6 GeV as shown in Fig. 1.

The SciBooNE detector consists of three detector components; SciBar, Electromagnetic Calorimeter (EC) and Muon Range Detector (MRD). SciBar is a fully active and fine grained scintillator detector that consists of 14,336 bars arranged in vertical and horizontal planes. SciBar is capable of detecting all charged particles and performing $dE/dx$-based particle identification. The EC is located downstream of SciBar. The detector is a "spaghetti" calorimeter with thickness of $11X_0$, and is used to measure $\pi^0$ and the intrinsic $\nu_e$ component of the neutrino beam. The MRD is located downstream of the EC in order to measure the momentum of muons up to 1.2 GeV/$c$ with range. It consists of 2-inch thick iron plates sandwiched between layers of plastic scintillator planes.

## SEARCH FOR $\bar{\nu}$ CHARGED CURRENT COHERENT PION PRODUCTION

### Antineutrino beam and wrong-sign background

Figure 1 shows the neutrino flux predictions [24] at the SciBooNE detector location in antineutrino mode running, as a function of neutrino energy. As shown in the figure, a relatively large fraction of neutrino contamination is expected in antineutrino mode running; $\sim 84\%$ is muon-antineutrino, $\sim 16\%$ muon-neutrino (cf. $\sim 93\%$ pure muon-neutrino beam in neutrino running mode). The beam simulation and the NEUT generator [25] predict about 30% contamination from neutrinos in the total event rate (flux times cross section). Having precise knowledge of neutrino ("wrong-sign") backgrounds in data collected in antineutrino mode is important for any antineutrino cross section measurements.

Because SciBooNE does not have a magnetized detector, we cannot distinguish neutrino and antineutrino interactions by the charge of the final state muons. However, event topology in antineutrino interactions differ from neutrino's. SciBar is uniquely suited to provide a measurement of the wrong-sign contamination in the antineutrino beam. We exploit the fact that the fine-grained SciBar tracking detector can differentiate between final states with protons versus neutrons. Hence we can distinguish neutrino versus antineutrino charged current quasi elastic (CC-QE) interactions on an event-by-event basis: $\nu_\mu + n \rightarrow \mu^- + p$ and $\bar{\nu}_\mu + p \rightarrow \mu^+ + n$. This is done based on their differing final state composition: CC-QE neutrino interactions are expected to have two tracks (one each from the muon and proton) while antineutrino interactions are expected to have only one track (from the muon), since the neutron is not detected.

Figures 2 and 3 show the selected 1-track ($\mu$) events and 2-track $\mu + p$ events in the antineutrino mode sample. According to the MC simulation, the purity of right-sign ($\bar{\nu}$) in the 1-track sample is $\sim 80\%$, and wrong-sign fraction in 2-track $\mu + p$ sample is $\sim$ 70%. Both samples shows fairly good agreement between data and MC simulation, which means the flux prediction provides a reasonable understanding of wrong-sign contamination in antineutrino running mode.

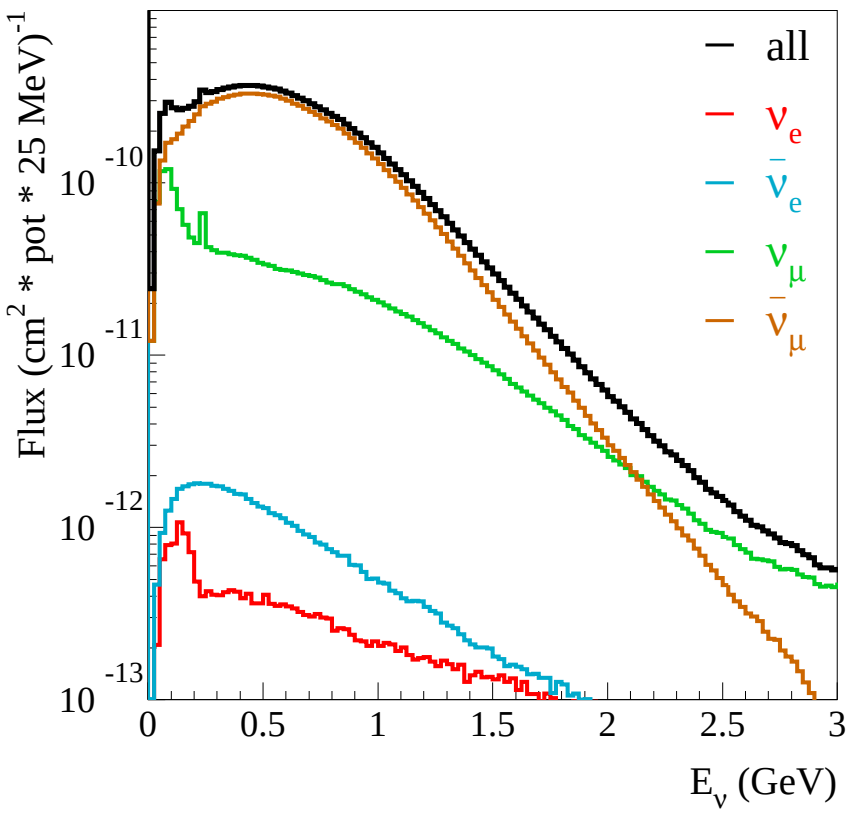

**FIGURE 1.** Antineutrino flux predictions at the SciBooNE detector, as a function of the neutrino energy. The total flux (black) is shown, as well as the flux per neutrino flavor: $\bar{\nu}_\mu$ (brown), $\nu_\mu$ (green), $\nu_e$ (red), $\bar{\nu}_e$ (blue). The averaged antineutrino energy is predicted to be $\sim$0.6 GeV.

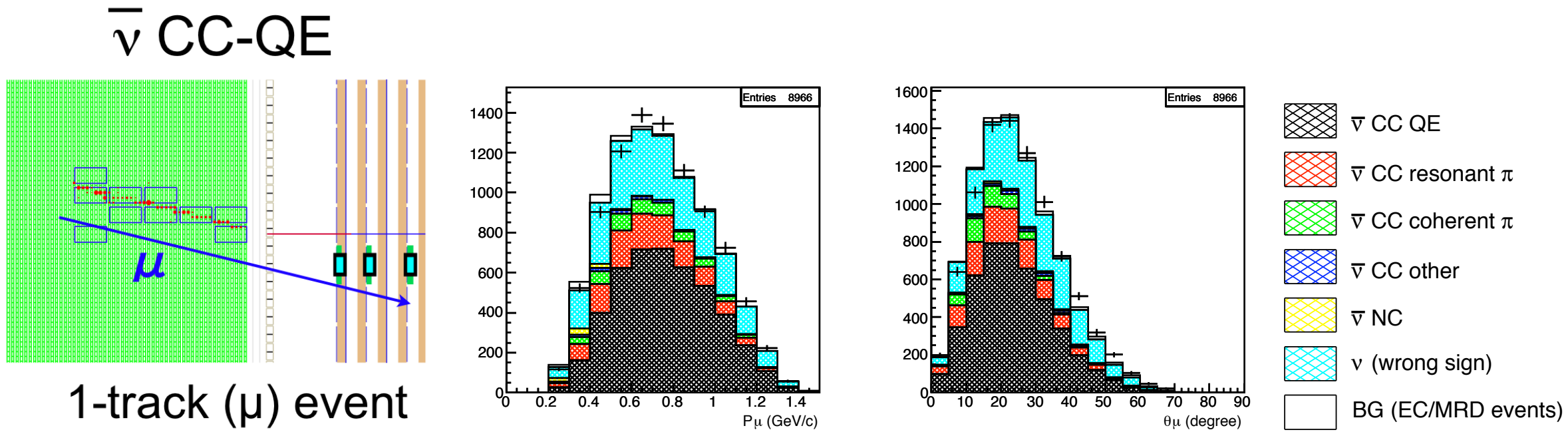

**FIGURE 2.** The selected 1-track sample in the antineutrino mode sample. Left panel shows a typical event display of 1-track ($\mu$) event. The right two plots show muon momentum (center) and muon angular (right) distributions of the sample. Data (with statistical error) are shown as crosses.

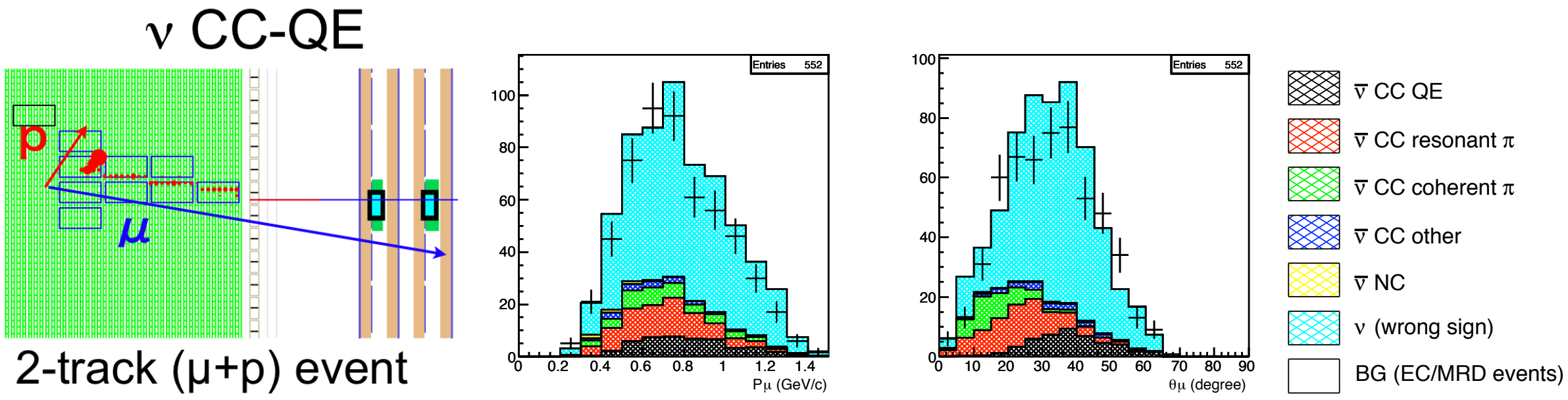

**FIGURE 3.** The selected 2-track $\mu + p$ sample in the antineutrino mode sample. The left panel shows a typical event display of 2-track $\mu + p$ event. The right two plots show muon momentum (center) and muon angular (right) distributions of the sample. Data (with statistical error) are shown as crosses.

## Event selection

The event selection criteria used in this antineutrino charged current coherent pion search are identical to those used in SciBooNE's neutrino charged current coherent pion search [14].

The experimental signature of charged current single charged pion production is the existence of two and only two tracks originating from a common vertex, both consistent with minimum ionizing particles (a muon and a charged pion). Even in the case of a charged current resonant pion event with a proton in the final state, $\nu_\mu n \rightarrow \mu^- p\pi^+$, the proton is often not reconstructed due to its low energy.

To identify charged current events, we search for tracks in SciBar matching with a track or hits in the MRD. Such a track is defined as a SciBar-MRD matched track. The most energetic SciBar-MRD matched track in any event is considered as the muon candidate. The matching criteria imposes a muon momentum threshold of 350 MeV/c. The neutrino interaction vertex is reconstructed as the upstream edge of the muon candidate. We select events whose vertices are in the SciBar fiducial volume (FV), 2.6 m$\times$2.6 m$\times$1.55 m, a total mass of 10.6 tons. Finally, event timing is required to be within a 2 $\mu$sec beam timing window.

This analysis uses events with the muon stopping in the MRD, classified as MRD stopped events. The average neutrino beam energy for true charged current events in the MRD stopped sample is $\sim$ 1.0 GeV.

Once the muon candidate and the neutrino interaction vertex are reconstructed, we search for other tracks originating from the vertex. Events with two and only two vertex-matched tracks are selected. The two track sample is divided based on particle identification. The SciBar detector has the capability to distinguish protons from muons and pions using $dE/dx$. The particle identification based on $dE/dx$ is applied to both tracks to select events with two MIP-like tracks ($\mu + \pi$). The probability of mis-identification is estimated to be 1% for muons and 10% for protons.

The sample still contains CC-QE, mainly wrong-sign CC-QE, events in which a proton is mis-identified as a minimum ionizing track. For each two-track event, we define an angle called $\Delta\theta_p$ as the angle between the observed second track direction and the expected proton track direction assuming a CC-QE event. Events with $\Delta\theta_p$ larger than 20 degrees are selected to further reduce the CC-QE contamination. Further selections are applied in order to isolate charged current coherent pion events from charged current resonant pion events which are the dominant backgrounds for this analysis. In the case of charged current coherent pion events, both the muon and pion tracks are directed forward. Events in which the track angle of the pion candidate with respect to the beam direction is less than 90 degrees are selected. To separate further charged current coherent pion events from charged current resonant pion events, additional protons with momentum below the tracking threshold are detected by their large energy deposition around the vertex, so-called vertex activity. We search for the maximum deposited energy in a strip around the vertex, an area of 12.5 cm$\times$12.5 cm in both views. Events with energy deposition greater than 10 MeV are considered to have activity at the vertex. In this paper, we define "charged current coherent pion sample" which do not have activity and its "counter-sample" that have activity.

Figure 4 and 5 show muon kinematics and reconstructed $Q^2$ distributions for the final selected charged current coherent sample and its counter-sample, respectively. Here, the reconstructed $Q^2$ is calculated as

$$Q^2_{rec} = 2E^{rec}_\nu (E_\mu - p_\mu \cos\theta_\mu) - m^2_\mu \qquad (1)$$

where $E^{rec}_\nu$ is the reconstructed neutrino energy calculated by assuming CC-QE kinematics,

$$E^{rec}_\nu = \frac{1}{2}\frac{(M^2_p - m^2_\mu) - (M_n - V)^2 + 2E_\mu(M_n - V)}{(M_n - V) - E_\mu + p_\mu \cos\theta_\mu} \qquad (2)$$

where $M_p$ and $M_n$ are the mass of proton and neutron, respectively, and $V$ is the nuclear potential, which is set to 27 MeV. Note that in this analysis a MC tuning to constrain the background events is not done yet.

As shown in Fig. 4 there is a data deficit in the low $Q^2_{rec}$ region, while the counter-sample, Fig 5, shows reasonable agreement between data and MC simulation. The data clearly lie above the MC predicted backgrounds in the coherent pion region below 0.1 (GeV/c)$^2$, although the predicted coherent pion signal is larger than what is observed in the data. These SciBooNE data suggest non-zero charged current coherent pion production, but it appears to be lower than that predicted by the Rein-Sehgal model employed by the NEUT generator. The statistical significance of the data excess is 4$\sigma$ below 0.1 (GeV/c)$^2$. It is interesting to note that the data excess above the predicted backgrounds are consistent with the SciBooNE (and K2K) upper limits observed in the neutrino mode search within the statistical uncertainties. Note that in this analysis the systematic uncertainties are not yet taken into account.

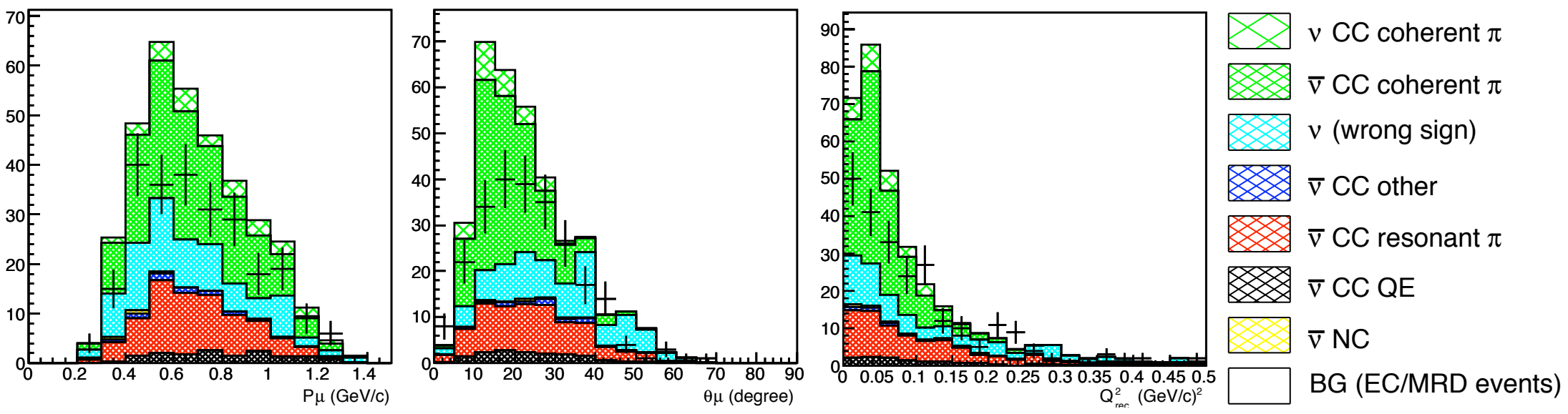

**FIGURE 4.** Antineutrino charged current coherent pion sample. The left two plots show muon momentum (left) and muon angular $\theta_\mu$ w.r.t. beam direction (center) and reconstructed $Q^2$ distributions of the sample.

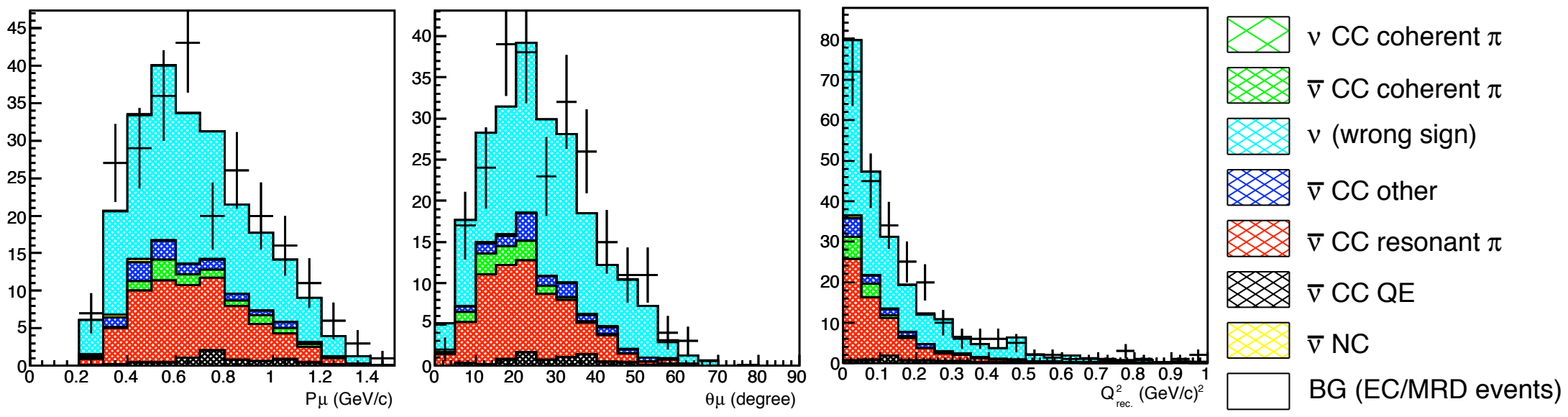

**FIGURE 5.** The counter-sample to the charged current coherent pion sample (Fig. 4). The plots show muon momentum (left) and muon angular (center) and reconstructed $Q^2$ distributions of the sample.

## DISCUSSION

We have investigated the data excess further by examining other kinematic distributions. Figure 6 shows pion kinematics in the antineutrino charged current coherent pion sample (the same sample shown in Fig. 4). As shown in the 2-dimensional histogram in $\theta_\mu$ and $\theta_\pi$, where $\theta_{\mu/\pi}$ are angles of muon and pion with respect to the beam direction, the data excess concentrates in small $\theta_\pi$ and $\theta_\mu$ region, in other words, the data excess is mainly visible at smaller opening angles between muon and pion tracks.

Figure 7 shows the reconstructed event kinematics, $E_\nu^{rec}$ and $Q^2_{rec}$, distributions of two distinct subsamples of antineutrino charged current coherent pion sample for poin scattered angles less than 35 degrees and more than 35 degrees. As shown in Fig. 7, data in the sample $\theta_\pi > 35°$ are consistent with background prediction while there is an enhancement of the data excess in the sample $\theta_\pi \leq 35°$. The same test have been performed in SciBooNE's neutrino charged current coherent pion sample [26], and we found a similar data excess to that seen in antineutrino mode. SciBooNE's antineutrino and neutrino data suggest that pions from charged current coherent pion events tend to be produced more forward than the Rein-Sehgal model employed by the NEUT generator. In order to study charged current coherent pion production, pion kinematics is important as well as muon kinematics.

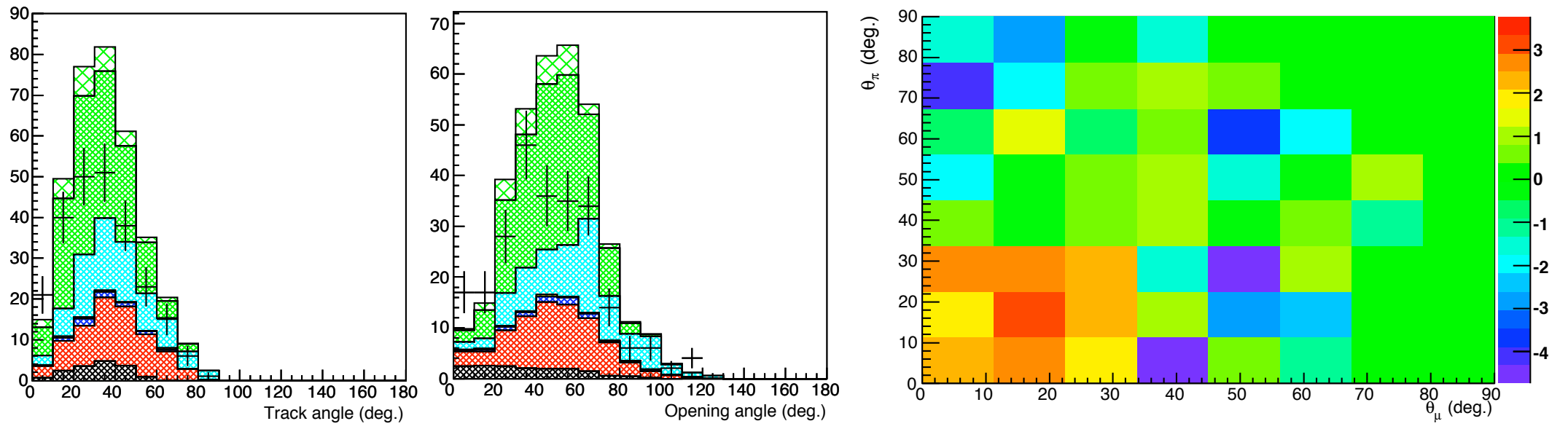

**FIGURE 6.** Pion kinematics distributions in antineutrino charged current coherent pion sample. Left plot shows pion angle $\theta_\pi$ with respect to neutrino beam direction, plot in center shows opening angle between muon track and pion track. Right plot is a 2-dimensional (2D) histogram in $\theta_\pi$ vs. $\theta_\mu$. The 2D histogram shows the difference between data and MC simulation and divided by statistical uncertainties. The MC simulation in the 2D histogram does not include charged current coherent pion prediction so that the plot shows the data excess with respect to the background prediction of NEUT generator. For example, the color of orange to red in the 2D histogram corresponds to 2- to 3-$\sigma$ data excess.

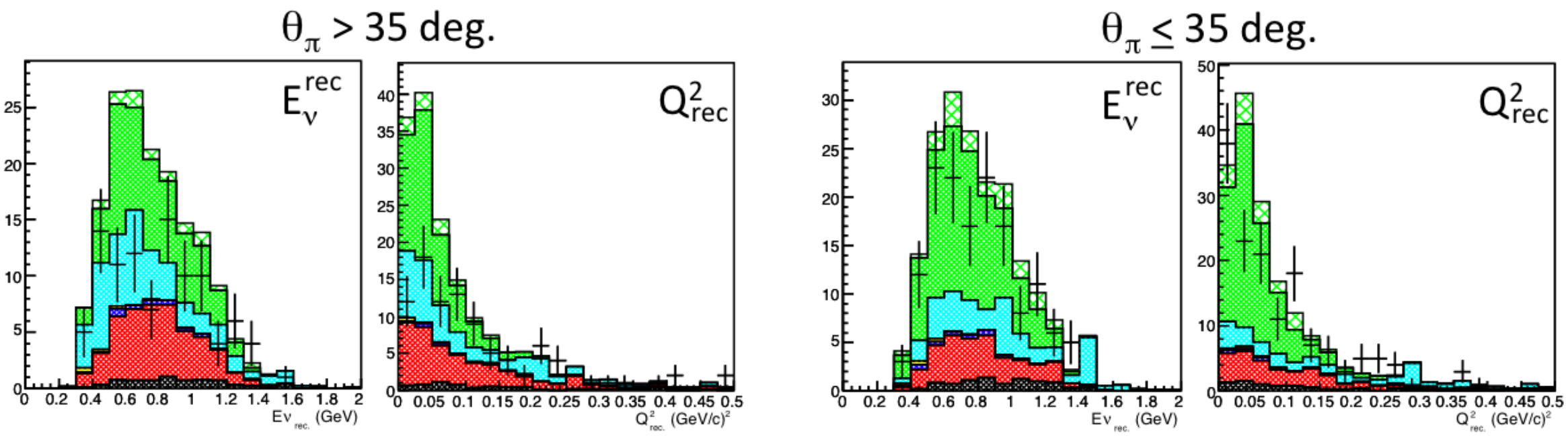

**FIGURE 7.** Left two plots shows reconstructed neutrino energy and $Q^2$ distributions for $\theta_\pi > 35$ deg. Right two plots shows reconstructed neutrino energy and $Q^2$ distributions in $\theta_\pi < 35$ deg.

## ACKNOWLEDGMENTS

The author would like to acknowledge support from the JSPS Postdoctoral Fellowship for Research Abroad, Columbia University and Massachusetts Institute of Technology. The SciBooNE Collaboration gratefully acknowledge support from various grants, contracts and fellowships from the MEXT (Japan), the INFN (Italy), the Ministry of Education and Science and CSIC (Spain), the STFC (UK), and the DOE and NSF (USA).